\documentclass[%
aps,
prd,
twocolumn,
preprintnumbers,
nofootinbib,
amsmath,amssymb,
]{revtex4-2}

\usepackage{style}
\usepackage{aas_macros}
\newcommand{\holodeck}{\texttt{holodeck}}

\begin{document}

\preprint{DESY-24-097}

%\title{GWB Anisotropies Forecasts}
\title{Detecting Gravitational Wave Anisotropies from Supermassive Black Hole Binaries}

\author{Anna-Malin Lemke\,\orcidlink{0009-0005-3568-3336}}
\email{anna-malin.lemke@desy.de}
\affiliation{II. Institute of Theoretical Physics, Universität Hamburg, Luruper Chaussee 149, 22761, Hamburg, Germany}

\author{Andrea Mitridate\,\orcidlink{0000-0003-2898-5844}}
\email{andrea.mitridate@desy.de}
\affiliation{Deutsches Elektronen-Synchrotron DESY, Notkestr. 85, 22607 Hamburg, Germany}

\author{Kyle A. Gersbach\,\orcidlink{0000-0003-2898-5844}}
\email{kyle.gersbach@nanograv.org}
\affiliation{Department of Physics and Astronomy, Vanderbilt University, 2301 Vanderbilt Place, Nashville, Tennessee 37235, USA}

\begin{abstract}
Several Pulsar Timing Array (PTA) collaborations have recently found evidence for a gravitational wave background (GWB) permeating our galaxy. The origin of this background is still unknown. Indeed, while the gravitational wave emission from inspiraling supermassive black hole binaries (SMBHB) is the primary candidate for its origin, several cosmological sources have also been proposed. One distinctive feature of SMBHB-generated backgrounds is the presence of GWB anisotropies stemming from the binaries distribution in the local Universe. However, none of the anisotropy searches performed to date reported a detection. In this work, we show that the lack of anisotropy detection is not currently in tension with a SMBHB origin of the background.
We accomplish this by calculating the anisotropy detection probability of present and future PTAs. We find that a PTA with the noise characteristics of the NANOGrav 15-year data set had only a $~2\%-11\%$ probability of detecting SMBHB-generated anisotropies, depending on the properties of the SMBHB population. However, we estimate that for the IPTA DR3 data set these probabilities will increase to $~4\%-28\%$, putting more pressure on the SMBHB interpretation in case of a null detection. We also identify SMBHB populations that are more likely to produce detectable levels of anisotropies. This information could be used together with the spectral properties of the GWB to characterize the SMBHB population.
\end{abstract}

\maketitle
\tableofcontents

\section{Introduction}
Several Pulsar Timing Array (PTA) collaborations have recently found strong evidence for the presence of a gravitational wave background (GWB) in the nHz frequency band~\cite{ng+23_gwb, EPTA:2023fyk, Reardon:2023gzh}. The most plausible source for such a background is a population of supermassive black hole binaries (SMBHB) formed in galaxy mergers through cosmic history. However, cosmological sources (including cosmic inflation, scalar-induced GW, phase transitions, cosmic strings, and domain walls) have also been proposed as a possible source of the background (see, for example, Ref.~\cite{NANOGrav:2023hvm}). Discriminating between these two possible interpretations is one of the main near-term goals of PTA experiments.

Searches for anisotropies in the GWB provide a promising path to establishing the origin of the GWB. Indeed, if the GWB is produced by a population of inspiraling SMBHB, clustering of host galaxies and Poisson fluctuations in binaries properties are expected to induce a detectable level of anisotropies in the GWB~\cite{Mingarelli:2017fbe}. On the other hand, the level of anisotropies produced by most cosmological sources is well below the sensitivity of present and future PTAs~\cite{Caprini:2018mtu, LISACosmologyWorkingGroup:2022kbp}. Therefore, detecting anisotropies in the GWB would provide strong evidence in favor of an astrophysical origin of the signal. At the same time, we might use the lack of GWB anisotropies to constrain the properties of the SMBHB population and eventually rule out the astrophysical interpretation in favor of a cosmological one.

The recent search for anisotropies in NANOGrav 15-year data set has reported null evidence, and placed an upper limit on the level of broadband anisotropies~\cite{ng+23_anis}. In view of these results a few questions arise: Is the lack of evidence for GWB anisotropies in tension with an SMBHB origin of the background? What can we learn about the SMBHB population given this null detection? Are the properties of the SMBHB population needed to reproduce the intensity of the observed signal compatible with a null anisotropy detection?

In this work, we address these questions by estimating the anisotropy detection probability for GWBs sourced by SMBHB populations whose demographic and evolution are consistent with the background spectral properties. We do this by running realistic anisotropy searches on mock PTA data in which we injected the GW signal generated from synthetic SMBHB populations produced using \texttt{holodeck}~\cite{holodeck}. Specifically, we proceed as follows: for a given set of SMBHB population parameters, we generate several mock SMBHB populations and their associated PTA signal, specifically the pulsar-pair correlations that such populations would induce (see Sec.~\ref{sec:mock_data}). We then run a frequentist anisotropy search~\cite{Pol:2022sjn} on these mock pulsar-pairs correlations, and derive the anisotropy detection probability for a given set of population parameters (see Sec.~\ref{sec:search_strat} and Sec.~\ref{sec:det_stat} for details). Finally, we discuss our results (see Sec.~\ref{sec:results}), which consist in a mapping between sets of SMBHB population parameters and anisotropies detection probabilities. This mapping gives us the tools to interpret the results of present and future searches for anisotropies. Specifically, it allows us to assess if a null anisotropy detection can be considered in tension with an SMBHB origin of the background, and use the results of anisotropy searches to better characterize the properties of the SMBHB population.

\section{How to Search for Anisotropies}\label{sec:search_strat}
In this section, we provide a summary of how GWB anisotropies appear in PTA data and discuss some of the data analysis techniques commonly adopted to extract these signals. 
\subsection{Anisotropies and pulsar-pair correlations}
PTAs track the arrival times of radio pulses emitted by a collection of galactic millisecond pulsars. The presence of a GWB will appear in PTA data as correlated perturbations, $\delta t$, in the arrival times of these pulses. Specifically, let's consider a GWB whose plane-wave expansion is given by
\begin{equation}
    h_{ij}(t,\vec{x})=\sum_A\int df\int_{S^2}d\hat\Omega\; h_A(f,\hat\Omega) e^{i 2\pi f(t-\hat\Omega\cdot\vec{x})}e_{ij}^A(\hat{\Omega})\,,
\end{equation}
where $f$ is the GW frequency, $A=+,\times$ labels the two GW polarizations, $e_{ij}^A$ are the GW polarization tensors, and $\hat\Omega$ identifies the propagation direction for a single GW plane wave. For a stationary, Gaussian, and unpolarized background the polarization amplitudes, $h_A(f,\hat{\Omega})$, are complex random variables with vanishing expectation values and a two-point correlator given by:
\begin{equation}
    \langle h_A^*(f,\hat\Omega)h_{A'}(f',\hat\Omega')\rangle = \frac{1}{16\pi}\delta_{AA'}\delta(f-f')\delta(\hat{\Omega},\hat{\Omega}')S_h(\hat\Omega, f)\,.
\end{equation}
The GWB power spectrum can be factorized as $S_h(\hat{\Omega},f)=S_h(f)P(\hat{\Omega},f)$, where the function $S_h(f)$ describes the spectral content of the GWB, and $P(\hat{\Omega},f)$ describes the distribution of the GWB power in the sky. 

The expected correlations between the perturbations in the pulse time of arrivals, usually referred to as timing residuals, for pulsar $a$ at time $t_i$ and the timing residials for pulsar $b$ at time $t_j$ induced by such a GWB are given by~\cite{Allen:1997ad}:
\begin{equation}
    \langle \delta t_a(t_i)\delta t_b(t_j)\rangle=\int \frac{df}{24\pi^2 f^2}\,e^{2\pi if(t_i-t_j)}\,S_h(f)\,\Gamma_{ab}(f)\,.
\end{equation}
From this equation is clear that $S_h(f)$ encodes the time correlation of the residuals. The spatial correlations of the residuals are instead encoded in the overlap reduction function, $\Gamma_{ab}$, which is related to the sky-distribution of the GWB power as:\footnote{Here we are neglecting the pulsar term contribution to the overlap reduction function. While this is the standard assumption in all anisotropy searches, an in-depth analysis of how this assumption affects anisotropy searches will be presented in a follow-up work~\cite{konstandin:2024}.}
\begin{equation}\label{eq:theo_corr}
    \Gamma_{ab}(f)=\frac{3}{2}\int_{S^2}\frac{d\hat\Omega}{4\pi}\,P(\hat\Omega,f)\sum_{A=+,\times}\mathcal{F}^A(\hat p_a,\hat\Omega)\mathcal{F}^A(\hat p_b,\hat\Omega)
\end{equation}
where $\mathcal{F}^A(\hat p,\hat\Omega)$ is the antenna response function for the $A$-th GW polarization:
\begin{equation}
    \mathcal{F}^A(\hat p,\hat\Omega)=\frac{\hat p^i\hat p^j}{2(1-\hat\Omega\cdot\hat p)}e_{ij}^A(\hat\Omega)\,.
\end{equation}

Working with a discrete set of frequencies, indexed by $n$, and rewriting the integral in Eq.~\eqref{eq:theo_corr} as a sum over equal-areal pixels, indexed by $k$, we get:
\begin{equation}\label{eq:theo_corr_disc}
    \Gamma_{ab,n}=\frac{3}{2}\sum_k P_{n,k}\left[\mathcal{F}_{a,k}^+\mathcal{F}_{b,k}^++\mathcal{F}_{a,k}^\times\mathcal{F}_{b,k}^\times\right]\frac{\Delta\hat\Omega_k}{4\pi}\,.
\end{equation}
In this work, we will normalize the GWB power such that $\int d\hat\Omega\,P(\hat\Omega,f_n)=\sum_k P_{n,k}\Delta\hat\Omega_k=4\pi$. With this normalization, a perfectly isotropic GWB is given by $P_{n,k}=1$, and when plugged into Eq.~\eqref{eq:theo_corr_disc} it returns the familiar Hellings-Downs correlation pattern~\cite{Hellings:1983fr}.

\subsection{Anisotropy searches}
Given a set of timing residuals, we can estimate the cross-correlation coefficients using the frequentist estimator developed in Refs.~\cite{Demorest:2012bv, Siemens:2013zla, Chamberlin:2014ria, Vigeland:2018ipb}:
\begin{equation}\label{eq:os_corr}
    \rho_{ab}=\frac{\delta\mathbf{t}_a^T\mathbf{N}_a^{-1}\hat{\mathbf{S}}_{ab}\mathbf{N}_b^{-1}\delta\mathbf{t}_b^T}{\tr\left[\mathbf{N}_a^{-1}\hat{\mathbf{S}}_{ab}\mathbf{N}_b^{-1}\hat{\mathbf{S}}_{ba}\right]},
\end{equation}
where $\rho_{ab}$ are the best estimates for the cross-correlation coefficients, $\delta\mathbf{t}_a$ contains the timing residuals for pulsar $a$, $\mathbf{N}_{a}=\langle\delta\mathbf{t}_a\delta\mathbf{t}_a^T\rangle$ is the $a$-th pulsar's autocovariance matrix, and $\hat{\mathbf{S}}_{ab}$ is the template-scaled covariance matrix. This template-scaled covariance matrix encodes information about the spectral shape of the GWB but is agnostic about its overall amplitude and cross-correlation signature. It is related to the full covariance matrix, $\mathbf{S}_{ab}\equiv\langle\delta\mathbf{t}_a\delta\mathbf{t}_b^T\rangle$, as $\mathbf{S}_{ab}=A^2 \Gamma_{ab}\hat{\mathbf{S}}_{ab}$, where $A$ is the GWB amplitude for a given spectral shape template. 
The values of $\mathbf{N}_a$ and $\hat{\mathbf{S}}$ are not known a priori, and need to be extracted from the data, usually using Bayesian inference techniques.

Equation \eqref{eq:os_corr} allows us to estimate broad-band correlations for the timing residuals. However, the GWB sky sourced by realistic SMBHB populations will not be constant across frequencies, and therefore induce correlations which are not constant in frequency space. These frequency-resolved correlations can be estimated using the recently developed per-frequency Optimal Statistic (PFOS) methods described in Ref.~\cite{pfos}. Assuming stationary Gaussian noise for the cross-correlation uncertainties, the likelihood function for these frequency-resolved correlations, $\vec{\rho}_n$, given a GWB sky, $\mathbf{P}_n$, can be written as:\footnote{Generally, cosmic variance and deviations from isotropy will source off-diagonal elements in the noise covariance matrix~\cite{Allen:2022ksj}. However, since the scope of this work is to closely replicate  current anisotropy searches, we ignore these off-diagonal elements following the same convention adopted in Ref.~\cite{ng+23_anis}.}
\begin{equation}
    p(\boldsymbol{\rho}_n\vert\mathbf{P}_n)=\frac{\exp[-\frac{1}{2}(\boldsymbol{\rho}_n-\mathbf{R}\mathbf{P}_n)\mathbf{\Sigma}^{-1}_n(\boldsymbol{\rho}-\mathbf{RP}_n)]}{\sqrt{\det(2\pi\mathbf{\Sigma}_n)}},
    \label{eq:likelikood}
\end{equation}
where $\boldsymbol{\Sigma}_n$ is the $N_\text{cc}\times N_\text{cc}$ diagonal noise covariance matrix for the cross-correlation measurements in the $n$-th frequency bin, and we have defined the PTA overlap response matrix
\begin{equation}    
    R_{ab,k} = \frac{3}{2N_{pix}}[\mathcal{F}_{a,k}^{+}\mathcal{F}_{b,k}^{+} + \mathcal{F}_{a,k}^{\times}\mathcal{F}_{a,k}^{\times}]
    \label{eq:or_matrix}.
\end{equation}
Therefore, given a set of measured cross-correlations, we can maximize this likelihood to obtain an estimate of the GWB sky, $\hat{\mathbf{P}}_n$. In this work, we choose to work in the pixel basis~\cite{Cornish:2014rva}, in which the GWB sky is divided into independent equal-area pixels. In the pixel basis, reconstructing the GWB sky corresponds to reconstructing the GWB power in each of these equal area pixels. Other typical basis choices include decomposition of the sky into linear~\cite{Mingarelli:2013dsa, Gair:2014rwa, Taylor:2013esa, Taylor:2015udp} and square root spherical harmonics~\cite{Taylor:2020zpk} or eigenmaps of the PTA Fisher matrix~\cite{Ali-Haimoud:2020iyz}. 

In the rest of this section, we discuss the three techniques used in this work to maximize the likelihood function in Eq.~\eqref{eq:likelikood}. Examples of sky maps reconstructed by using some of these techniques are shown in Fig.~\ref{fig:mock_sky}.

\subsubsection{Analytical Solution}
In the pixel basis, there exists an analytical maximum solution for the likelihood in Eq.~\eqref{eq:likelikood}, given by~\cite{Thrane:2009fp,Romano:2016dpx,Ivezic_2013}:
\begin{equation}
    \mathbf{\hat{P}}_n=\mathbf{M}_n^{-1}\mathbf{X}_n,
    \label{eq:ML_solution}
\end{equation}
where we have defined the Fisher information matrix, $\mathbf{M}_n=\mathbf{R}^{T}\mathbf{\Sigma}^{-1}_n\mathbf{R}$, and the ``dirty map", $\mathbf{X}_n=\mathbf{R}^{T}\mathbf{\Sigma}^{-1}_n\boldsymbol{\rho}_n$. The diagonal elements of the Fisher matrix provide an estimate of the uncertainty on the recovered pixel values, while the dirty map provides an inverse noise-weighted representation of the power as seen by the PTA.

The inversion of the Fisher matrix typically limits the resolution of our maximum likelihood sky maps beyond the fundamental PTA limit. The spatial resolution of a PTA is approximately set by requiring $N_\text{pix}\leq N_\text{cc}$~\cite{Romano:2016dpx}, where $N_{cc}$ is the number of cross-correlations in the data set, and $N_{\rm pix}$ is the number of pixels in the sky-map tessellation. For a HEALpix~\cite{healpix} sky map, $N_{\rm pix}$ is related to the $N_\text{side}$ parameter as $N_\text{pix}=12\cdot N_\text{side}$. Given the NANOGrav 15yr configuration with 67 pulsars and $N_\text{cc}=2211$, this translates to $N_\text{side}\leq8$. However, we find the inverse of the Fisher matrix to be numerically ill-defined for $N_\text{side}=8$, which limits the resolution of the NANOGrav 15-year data set to $N_\textrm{side}\leq4$, corresponding to 12 ($N_\textrm{side}=1)$, 48 ($N_\text{side}=2$) and 192 ($N_\text{side}=4$) pixels for the sky maps.

Another downside of this map-reconstruction method is that it usually produces maps with negative power in some regions of the sky. These negative-power regions are --of course-- unphysical and prevent us from normalizing the recovered maps.

\subsubsection{Radiometer Search}
If we assume that most of the GW power is coming from a single bright pixel, instead of trying to reconstruct the entire GW sky, we can focus on reconstructing the power in each pixel independently, neglecting correlations between neighbouring pixels. The power in each pixel can be derived by replacing the Fisher matrix with its diagonal components in Eq.~\eqref{eq:ML_solution}. 

This reconstruction method, usually known as \emph{radiometer}~\cite{Ballmer:2005uw, Mitra:2007mc}, is best suited for GWB skies that contain a single, very bright hot spot instead of a collection of sources with similar signal amplitudes. While some SMBHB populations might generate GW skies with very bright hot spots, it is not clear a priori if a radiometer search is best suited to recover generic SMBHB skies. As part of our analysis, we will assess how well the radiometer search performs on realistic SMBHB skies.

\subsubsection{Numerical Solution}
Finally, we can maximize the cross-correlation likelihood numerically. Specifically, we can reconstruct the GW sky by numerically solving the following equation\footnote{We use the CVXPY package~\cite{diamond2016cvxpy, agrawal2018rewriting} to obtain the least-squares solution to Eq.~\eqref{eq:num_solution} while imposing positivity and the map normalization $\sum_{k}P_{n,k}\Delta\hat\Omega_k=4\pi$ as boundary conditions.}
\begin{equation}\label{eq:num_solution}
    \mathbf{M}_n\mathbf{\hat{P}}_n = \mathbf{X}_n\,.
\end{equation}
The advantage of this method is that it allows us to impose positivity on the recovered sky maps and therefore interpret each pixel value as physical power. Additionally, this approach does not rely on our ability to calculate the inverse of $\mathbf{M}_n$. 

\section{Mock Data}\label{sec:mock_data}
In this section we describe the procedure we used to generate mock PTA data. While, in principle, we could produce a set of mock timing residuals for each synthetic SMBHB population and then derive the corresponding pulsar cross-correlations, we instead directly generate the cross-correlation coefficients, $\rho_{ab,n}$, to maximize the computational efficiency of our procedure. Specifically, we proceed by following these three main steps:

\subparagraph{Population Synthesis (Sec.~\ref{subsec:smbhb_gen}) --}\hspace{-10pt} We explore the parameter space describing formation and evolution of SMBHBs. For each point in this space, we generate several realizations of the cosmic SMBHB population using the \texttt{holodeck} package~\cite{holodeck}.

\subparagraph{Sky Creation (Sec.~\ref{subsec:sky_gen}) --}\hspace{-10pt} For each SMBHB population generated in the previous step, we create a GWB sky, $P_{n,k}$, by randomly placing the population binaries in the sky. 

\subparagraph{Correlation Creation (Sec.~\ref{subsec:cor_gen}) --}\hspace{-10pt} For each GWB sky, we generate theoretical pulsar-pair correlation according to Eq.~\eqref{eq:theo_corr_disc}. To simulate measurement noise, on top of these theoretical correlations, we add zero-mean Gaussian noise whose standard deviation is given by the cross-correlations uncertainty measured from the NANOGrav 15-year data set~\cite{ng+23_anis}.

\subsection{Population Synthesis with \holodeck}\label{subsec:smbhb_gen}
To generate SMBHB populations, we used the same procedure as in Ref.~\cite{ng+23_astro}, which we briefly summarize in this section. For more details, we refer the reader to Refs.~\cite{ng+23_astro, holodeck}.
\renewcommand{\arraystretch}{1.25}
\setlength{\tabcolsep}{0.3cm}
\begin{table}[t!]
    % \centering
    \begin{tabular}{ccc}
    \toprule
         {\bf Model Component} & {\bf Parameter} &  {\bf Fiducial value} \\
        \hline
            \multirow{ 4}{*}{GSMF~\cite{Chen:2018znx}}    & $\psi_z$   & -0.60 \\
                    & $m_{\psi,z}$   & +0.11 \\
                    & $\alpha_{\psi,0}$   & -1.21 \\
                    & $\alpha_{\psi,z}$   & -0.03 \\

        \hline
         \multirow{2}{*}{$M_{\text{BH}}-M_\text{bulge}$~\cite{Gultekin:2009qn, Kormendy:2013dxa, McConnell:2012hz,Bluck:2014ila, Lang:2014cta}} & $\alpha_\mu$     & 1.10 \\
                    & $f_{\star,\text{bulge}}$     & 0.615 \\
        \hline
         \multirow{3}{*}{$\dfrac{da}{dt}$~\cite{Kelley:2016gse, ng+23_astro}}    & $a_c/{\rm pc}$  & $10^2$\\
                                        & $a_\text{init}/{\rm pc}$ & $10^4$\\
                                        & $\nu_\text{outer}$ & 2.5\\
    \bottomrule
    \end{tabular}
    \caption{Fiducial model parameters as presented in table B1 of Ref.~\cite{ng+23_astro}. }
    \label{tab:fid_val}
\end{table}
\subsubsection{Galaxy Merger Rate}
As SMBHB are formed in galaxy mergers, one of the key ingredients needed to model SMBHB populations is the rate at which such mergers take place. The comoving number density of galaxies mergers, $\eta_{\rm gal-gal}$, as function of the stellar mass of the primary galaxy, $m_{1\star}$, the stellar mass ratio, $q_\star\equiv m_{2\star}/m_{1\star}$, and redshift, $z$, can be written as~\cite{Chen:2018znx}:
\begin{equation}\label{eq:galaxy merger rate}
    \frac{\partial^{3}\eta_\text{gal-gal}}{\partial m_{\star 1}\partial q_\star \partial z}=\frac{\Psi(m_{\star 1},z^\prime)}{m_{\star 1}\ln(10)}\frac{P(m_{\star 1},q_\star,z^\prime)}{T_\text{gal-gal}(m_{\star 1},q_\star,z^\prime)}\frac{\partial t}{\partial z^\prime},
\end{equation}
where $\Psi$ denotes the galaxy stellar-mass function (GSMF), $P$ the galaxy pair fraction and $T_\text{gal-gal}$ the galaxy merger time. Since the galaxy merger spans a finite timescale ($T_{\rm gal-gal}$), in Eq.~\eqref{eq:galaxy merger rate}, we distinguish between the redshift at which the galaxy pair forms ($z' = z'[t]$ at some initial time $t$) and the redshift at which the system becomes a post-merger galaxy remnant ($z = z[t + T_{\rm gal-gal}]$).

The GSMF provides the differential number-density of galaxies per decade of stellar mass. Following Ref.~\cite{ng+23_astro}, we parametrize the GSMF as a single Schechter function~\cite{Schechter:1976iz}:
\begin{equation}
    \Psi(m_{\star1},z^{\prime}) = \ln(10)\Psi_{0}\cdot\bigg(\frac{m_{\star1}}{M_{\psi}}\bigg)^{\alpha_{\psi}}\exp\bigg(-\frac{m_{\star1}}{M_{\psi}}\bigg),
\end{equation}
where the phenomenological parameters $\Psi_0$, $M_\psi$ and $\alpha_\psi$ follow the redshift parametrization introduced in~\cite{Chen:2018znx}:
\begin{equation}
    \begin{split}
        \log_{10}(\Psi_0/\Mpc^{-3}) & = \psi_0 + \psi_z\cdot z\\
        \log_{10}(M_\psi/M_\odot) & = m_{\psi,0} + m_{\psi,z}\cdot z\\
        \alpha_\psi &= 1+\alpha_{\psi,0}+\alpha_{\psi,z}\cdot z
    \end{split}
\end{equation}
This parametrization introduces six free parameters for the GSMF. Of these six parameters, the mass normalization $\psi_0$ and the characteristic mass $m_{\psi,0}$ are allowed to vary in our analysis, while the remaining four are fixed to the fiducial values provided in Table~\ref{tab:fid_val}.

The galaxy pair fraction and merger time are kept fixed to fiducial power-law functions of the galaxy stellar mass, galaxy mass ratio, and initial redshift. For a full description of these functions we direct the reader to Ref.~\cite{ng+23_astro}.

\subsubsection{SMBH-Host Relation}
By populating galaxies with supermassive black holes (SMBHs), we can relate the galaxies merger rate in Eq.~\eqref{eq:galaxy merger rate} with the SMBHs merger rate. We populate the galaxies assuming a correspondence between the host galaxy and the SMBH at its center. Specifically, we assume that the SMBH mass, $M_{\rm BH}$, is related to the bulge mass of the host galaxy, $M_{\rm bulge}$, according to the relation~\cite{Marconi:2003hj}:
\begin{equation}
    \log_{10}(M_{\text{BH}}/M_{\odot})=\mu+\alpha_{\mu}\log_{10}\bigg(\frac{M_{\text{bulge}}}{10^{11}M_{\odot}}\bigg)+\mathcal{N}(0,\epsilon_\mu)\,
\end{equation}
where $\mathcal{N}(0,\epsilon_\mu)$ is a normally distributed random scatter with zero mean and standard deviation $\epsilon_\mu$, and $M_{\rm bulge}=f_{\star,{\rm bulge}}\cdot m_{\star}$ is the fraction of the galaxy stellar mass contained in the bulge. In our analysis, we vary the mass normalization $\mu$ and the scattering width $\epsilon_\mu$, while keeping the other parameters fixed to the fiducial values reported in Table~\ref{tab:fid_val}.

Using the $M_{\rm BH}-M_{\rm bulge}$ relation we can then translate the number density of galaxy mergers into a number density for SMBHB mergers, $\eta$, as:
\begin{equation}
    \frac{\partial^{3}\eta}{\partial M\partial q\partial z}=\frac{\partial^{3}\eta_\text{gal-gal}}{\partial m_{\star 1}\partial q_\star \partial z}\frac{\partial m_{\star 1}}{\partial M}\frac{\partial q_\star}{\partial q}\,,
\end{equation}
where $M=m_1+m_2$ is the total binary mass, $q\equiv m_2/m_1< 1$ is the binary mass ratio, and $m_{1,2}$ are the masses of the SMBHs in the binary system.
\renewcommand{\arraystretch}{1.45}
\setlength{\tabcolsep}{0.3cm}
\begin{table}[t!]
    % \centering
    \begin{tabular}{ccc}
    \toprule
         {\bf Binary property} & {\bf Range} &  {\bf Number of bins} \\
        \hline
        $M/M_\odot$ & $[10^4,10^{12}]$   & 91 \\
        $q$ & $[10^{-3},1]$   & 81 \\
        $z$ & $[10^{-3},10]$   & 101 \\
        $f/\rm{nHz}$ &  $[1.85, 5.93]$ & 3\\
    \bottomrule
    \end{tabular}
    \caption{Details of the SMBHB parameter space binning. Frequency bins are linearly spaced, while the bins for all the other dimensions are logarithmically spaced.}
    \label{tab:binning}
\end{table}

\subsubsection{Binary Evolution}
For binaries on circular orbits, the observer-frame GW frequency, $f$, is related to the binary's rest frame orbital frequency, $f_p$, as $f=2f_p/(1+z)$. The orbital frequency, in turn, is uniquely related to the binary orbital separation, $a$, according to $f_p^2=4\pi G M/a^3$.

The evolution of the binary system's orbital separation, $a$, is driven by a combination of GW emission and interactions with the astrophysical environment. The GW contribution to the binary hardening is given by~\cite{Peters:1964zz}:
\begin{equation}
    \frac{da}{dt}\bigg\vert_{\text{GW}}=-\frac{64G^3}{5c^5}\frac{m_1 m_2 M}{a^3}\,,
    \label{eq:hardening_gw}
\end{equation}
where $G$ is the gravitational constant. 
We describe environmental effects using a phenomenological model which aims to capture the overall effects of more complicated explicit binary evolution models, while introducing a small number of free parameters:
\begin{equation}
    \label{eq:hardening_pheno}
    \frac{da}{dt}\bigg\vert_{\text{phenom}}=H_{a}\bigg(\frac{a}{a_{c}}\bigg)^{1-\nu_\text{inner}}\bigg(1+\frac{a}{a_c}\bigg)^{\nu_\text{inner}-\nu_\text{outer}}.
\end{equation}
Here, the power law indices $\nu_\text{inner}$ and $\nu_\text{outer}$ are introduced, dividing the hardening process into a small-separation and a large-separation regime with different power-law behaviours and a turnover at a critical separation $a_c$. While $\nu_\text{inner}$ is allowed to vary in our model, $\nu_\text{outer}$ is fixed to a standard literature value of $+2.5$~\cite{Kelley:2016gse}. The total rate of binary evolution is obtained by summing the two contributions:
\begin{equation}
    \frac{da}{dt}= \frac{da}{dt}\bigg\vert_{\text{GW}} + \frac{da}{dt}\bigg\vert_{\text{phenom}}\,.
\end{equation}

The normalization of the phenomenological hardening rate, $H_{a}$, is determined by requiring that the binary lifetime, from the initial separation $a_{\rm init}$ to the innermost stable circular orbit $a_{\rm isco}=6GM/c^2$, matches a target value given by:
\begin{equation}\label{eq:hardening_time}
    \tau_{f}=\int_{a_\text{init}}^{a_\text{isco}}\bigg(\frac{da}{dt}\bigg)^{-1}da\,.
\end{equation}
The value of $\tau_f$ is a free parameter that we allow to vary in our analysis.

Given a model for the binary orbital evolution, we can write the number of SMBHB per given mass, mass-ratio, redshift, and log orbital frequency, as~\cite{Rajagopal:1994zj, Jaffe:2002rt, Sesana:2008mz}:
\begin{equation}\label{eq:num_den}
    \frac{\partial^4 N}{\partial M\partial q\partial z \partial \ln f_p} = \frac{\partial^{3}\eta}{\partial M\partial q \partial z}\cdot\tau(f_p)\cdot 4\pi c(1+z)d_c^2\,,
\end{equation}
where $d_c$ is the comoving distance of a source at redshift $z$, and $\tau(f_p)$ is the binary hardening timescale, which characterizes the amount of time a SMBHB spends in a given frequency bin:
\begin{equation}
    \tau(f_p)\equiv f_p/(df_p/dt)= -\frac{2a}{3}\frac{da}{dt}\,.
\end{equation}

\subsubsection{Population Synthesis}
Given a set of astrophysical parameters, $\vec\theta=\{\psi_0,\,m_{\psi,0},\,\mu,\,\epsilon_\mu, \,\tau_f,\,\nu_{\rm inner}\}$, we generate random SMBHB populations by drawing the number of binaries in a given $(M,\,q,\,z,\,f)$ bin from a Poisson distribution centered around the expectation value
\begin{equation}
    N(M,q,z,f_p)=\frac{\partial^4 N}{\partial M\partial q\partial z \partial \ln f_p}\Delta M\, \Delta q\, \Delta z\, \Delta\ln f_p,
\end{equation}
where $(\Delta M,\,\Delta q,\,\Delta z,\,\Delta f_p)$ are the bin sizes for each SMBHB parameter. Details about the binning of the SMBHB parameter space are reported in Tab.~\ref{tab:binning}.

To ensure that the SMBHB populations considered in our analysis are consistent with the spectral properties of the observed GWB, we sample the astrophysical parameters from the posterior distributions derived by fitting the SMBHB signal to the observed GWB spectrum. Specifically, we sample from the Monte Carlo chains obtained for the \emph{Phenom+Astro} model in Ref.~\cite{ng+23_astro}. For each $\vec{\theta}$ in these chains, we then generate 50 independent SMBHB populations. 

\subsection{GWB Sky Maps}\label{subsec:sky_gen}
Given a mock SMBHB population, the next step consists in generating an associated GW sky.
We start by assuming that each of the SMBHB is on a circular orbit, and assigning them the $M$, $q$, $z$, and $f$ values corresponding to their bin centers. We can then derive their sky and polarization-averaged GW amplitude as~\cite{Finn:2000sy}:
\begin{equation}\label{eq:binary_strain}
    h_s^{2}(f)=\frac{32}{5c^8}\frac{(G\mathcal{M})^{10/3}}{d_c^2}(2\pi f_p)^{4/3}\,,
\end{equation}
where $\mathcal{M}=M q^{3/5}/(1+q)^{6/5}$, and $f_p=f (1+z)/2$. This amplitude can then be translated into a characteristic strain as~\cite{Rosado:2015epa}:
\begin{equation}\label{eq:char_binary_strain}
    h_c^{2}(f)= h_s^{2}(f)\frac{f}{\Delta f}\,,
\end{equation}
where the frequency bin width is given by $\Delta f=1/T_\text{obs}$.
%, is introduced due to the finite frequency resolution, the spectral density $S_h(f)$ is spread over $\Delta f$, which introduces a normalization factor $1/\Delta f$. 

For each frequency bin, we then place the loudest 20 sources on random pixels of a sky-tessellation derived using HEALpix~\cite{healpix}, and add their squared characteristic strain to the corresponding pixels. The remaining binaries are added to the sky map as an isotropic background (equally divided among all pixels) with a characteristic strain given by
\begin{equation}
    h^{2}_{c,\scriptscriptstyle{\rm BG}}(f) = \sum_{M,q,z,f_p}N(M,q,z,f_p)h_c^2(f)\,,
\end{equation}
where the sum runs over all binaries expect the loudest 20.\footnote{To generate the mock sky maps we use $N_\text{side}=16$, resulting in a resolution of $12\cdot N_\text{side}^{2}=3072$ pixels. We have checked that the finite resolution effects on the induced pulsar-pair correlations are negligible for $N_\text{side}\geq16$. The anisotropy of the GWB is mainly captured by the loudest 10 sources in terms of spherical harmonic coefficients~\cite{Gardiner:2023zzr}, which we have confirmed also applies for our pixel basis.} The maps are then normalized such that $\sum_k P_{n,k}\Delta\hat\Omega_k=4\pi$, such that $P_{n}=1$ describes an isotropic sky and a theoretical upper limit of $P_{n,k,\rm max}=4\pi/\Delta\hat\Omega=N_{pix}$ is imposed for the power in each pixel, which corresponds to a singular GW source being present.

\begin{figure*}[htp!]
    \centering
    \includegraphics[width=\textwidth]{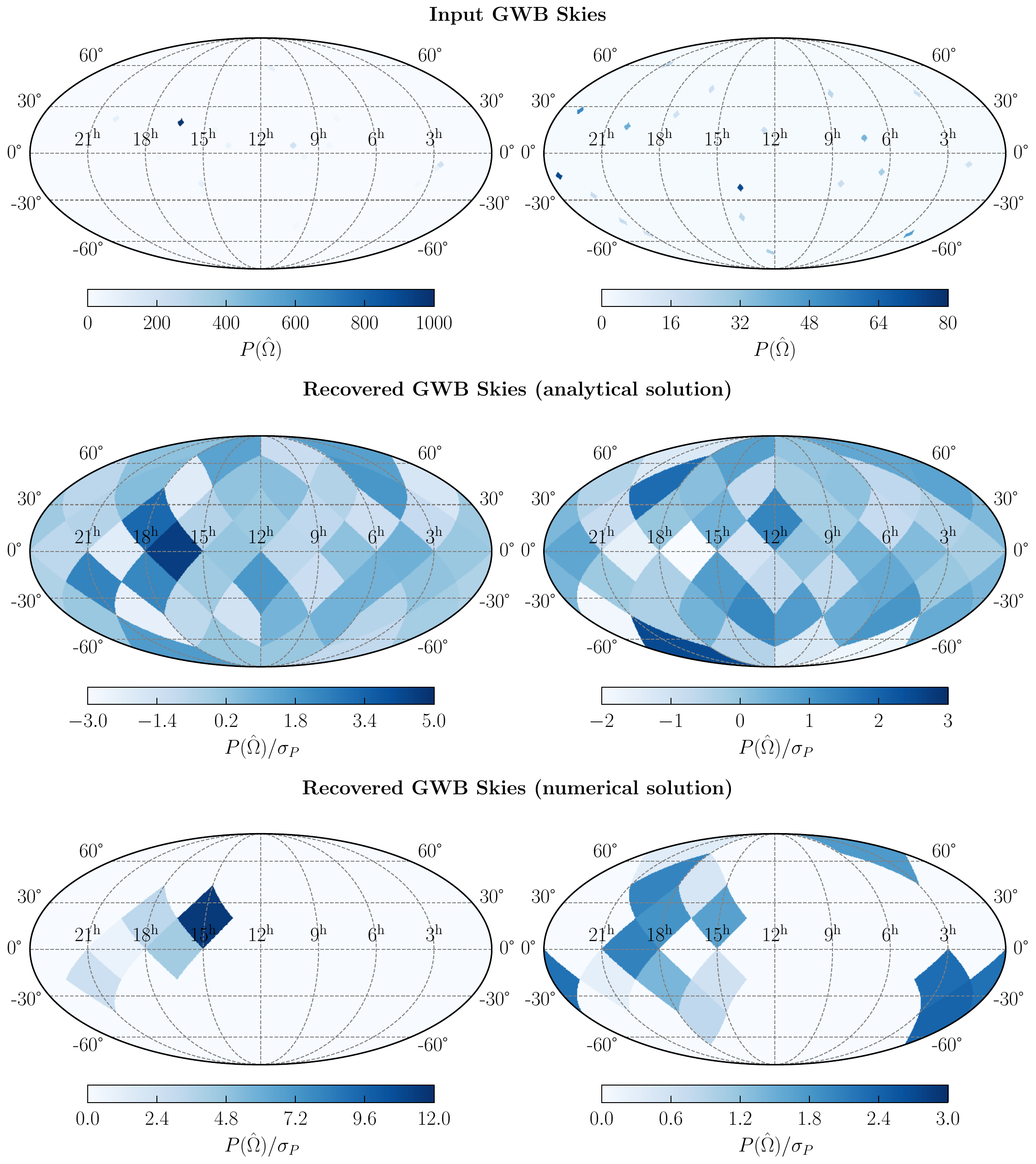}
    \caption{Examples of GWB power distributions induced by SMBHB populations (upper panel), and the reconstructed sky-maps obtained using the analytical (middle panel) and numerical (lower panel) reconstruction methods. The recovered pixel values are rescaled by their corresponding uncertainties derived from a null distribution of recovered sky maps from isotropic GW skies, $\sigma_P$, indicating the significance of deviation from isotropy. The left panels illustrate the reconstruction of a GW sky with a hot spot that leads to an anisotropy detection in our analysis. The right panels show a GW sky that does not produce a detection.}
    \label{fig:mock_sky}
\end{figure*}

To marginalize over the sky locations of the loudest sources, for each SMBHB population, we generate 30 independent sky maps per frequency bin. Given that we generate 50 SMBHB populations for each set of astrophysical parameters (see the previous section for details), we produce a total of 1500 sky maps per frequency bin for each set of astrophysical parameters. Examples of mock sky maps are provided in the upper panels of Fig.~\ref{fig:mock_sky}.

\begin{figure*}[ht!]
    \centering
    \includegraphics[width=\textwidth]{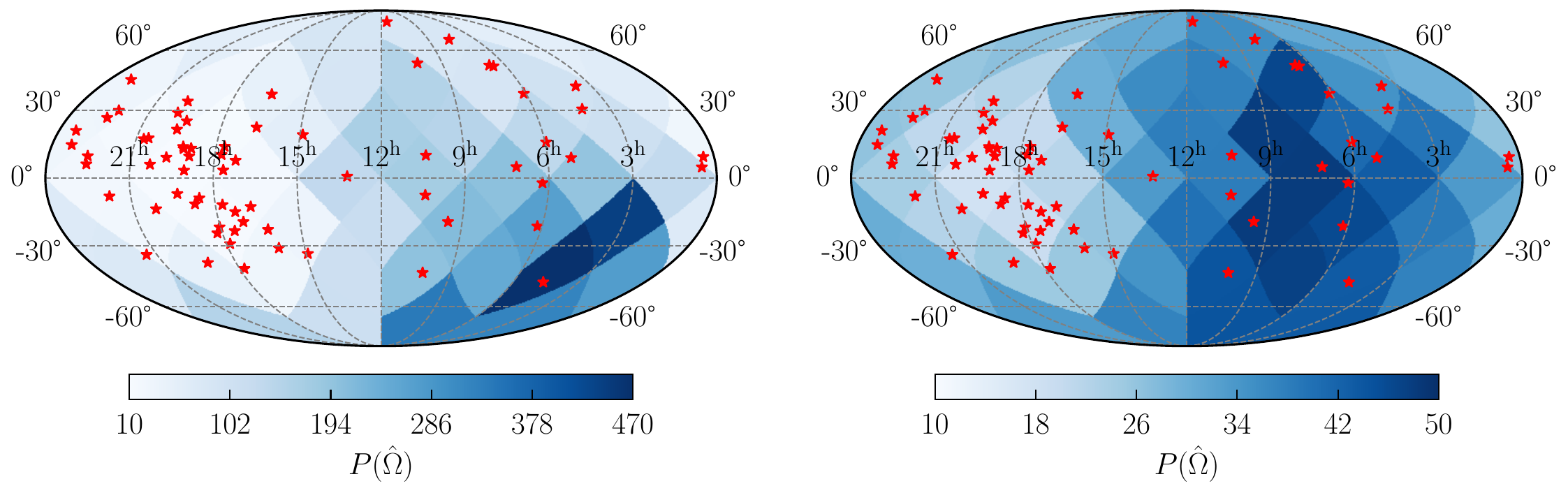}
    \caption{Pixel upper limits derived for $f_{\rm GW}=3.95$\nHz \ and $N_\text{side}=2$ using the analytical (left panel) and numerical (right panel) reconstruction methods. The red stars indicate the position of the pulsars contained in the NANOGrav 15-year data set. The higher pixel values for the analytical upper limits reflect our inability to normalize the reconstructed sky maps due to the presence of negative pixel values.}
    \label{fig:pixel_limits}
\end{figure*}

\subsection{Mock Correlations}\label{subsec:cor_gen}
The final step of the procedure consists of generating mock cross-correlations for each of the mock GWB skies. For a given sky map, $P_{i,k}$, the expected cross-correlations values are given by Eq.~\eqref{eq:theo_corr_disc}. However, any realistic measurement will not return these theoretical values due to finite experimental precision. Therefore, to simulate a realistic anisotropy search,  we generate mock correlation values by sampling from a Gaussian distribution centered around the expected value given by Eq.~\eqref{eq:theo_corr_disc}, and with a standard deviation given by the cross-correlation uncertainties measured from the NANOGrav 15-year data set using the PFOS~\cite{pfos}.

We use this procedure to generate two mock PTA data sets. The first data set consists of the cross-correlations values for the $67$ pulsars monitored for the NANOGrav 15-year data release. For the second data set, we try to mimic the capabilities of a future IPTA and generate mock cross-correlations for the 115 pulsars currently observed by NANOGrav~\cite{NANOGrav:2023hde} together with EPTA+InPTA~\cite{EPTA:2023sfo}, PPTA~\cite{PPTA}, and MeerKat~\cite{MeerKAT}.  Not having a realistic noise estimate for this data set, we use the cross-correlation uncertainties derived from the NANOGrav 15-year data set for the subset of pulsar pairs already included in this data set, and draw the noise values randomly from this set of cross-correlation uncertainties for pulsar pairs not included in the current NANOGrav observations.

It is important to note that, even for a PTA with infinite resolution, cosmic variance, polarization of the GW emission, and contributions from the pulsar term will induce deviations from the expected cross-correlations in Eq.~\eqref{eq:theo_corr_disc} (see, for example, Ref.~\cite{Allen:2022dzg}). While these deviations are below the current precision at which we can reconstruct the cross-correlations, their impact on our ability to reconstruct the underlying GWB sky has not yet been investigated. In this work, we will follow standard conventions of anisotropy searches, and ignore the impact of these effects. We leave a dedicated study of their impact to a follow-up analysis~\cite{konstandin:2024}.

\section{Detection Statistics}\label{sec:det_stat}
In the final step of our procedure, we take the mock cross-correlations generated for each SMBHB population and try to reconstruct the underlying GW sky by using the techniques discussed in Sec.~\ref{sec:search_strat}. For each of these reconstructed skies, we want to establish if they show any evidence for anisotropies. We do this by constructing an appropriate detection statistic that we then calibrate against a null distribution.

When searching for anisotropies, we assume isotropy as the null hypothesis. Therefore, we calibrate all our detection statistics against a null distribution derived by re-running our pipeline on mock cross-correlations generated from perfectly isotropic skies (instead of the SMBHB populated skies used in Sec.~\ref{sec:mock_data}). Specifically, we generate null distributions by proceeding as follows:
\begin{itemize}
    \item For each pulsar pair, we generate a cross-correlation value drawing from a normal distribution with a mean given by the Hellings \& Downs (HD) value and a standard deviation given by the cross-correlation uncertainties measured from the NANOGrav 15-year data set using the PFOS estimator~\cite{pfos}. We note that, even for isotropic skies, the cosmic variance effect already discussed in the previous section will induce deviations from HD correlations. In our analysis, following standard conventions (see for example Refs.~\cite{Mingarelli:2013dsa,Taylor:2015udp,ng+23_anis}), we ignore this effect. The impact of this assumption will be investigated in a follow-up analysis~\cite{konstandin:2024}.
    \item By using the methods discusses in Sec.~\ref{sec:search_strat}, we recover the GWB sky associated to these mock correlations.
    \item We compute the value of the detection statistic on the recovered GWB sky.
    \item We repeat this procedure $10^6$ times to derive a null distribution for the detection statistic.
\end{itemize}
By comparing the detection statistic values of a reconstructed sky with this null distribution we can assess the significance of the recovered anisotropies.

In the rest of this section, we discuss three detection statistics used in this work to define anisotropy detection. 

\subsection{Pixel upper limits}
An intuitive detection statistic is provided by the reconstructed GWB power in each pixel. 
By comparing the reconstructed pixel power with the pixel-specific null distribution, we can claim that a reconstructed sky map provides a detection with a $\sim3\sigma$ global significance if any of the reconstructed pixel power has a $p$-value smaller than $3\times 10^{-3}/N_{\rm pix}$. Here, 
the trial factor $1/N_{\rm pix}$ converts the local significance of each pixel to a global significance.\footnote{Here we are assuming that the recovered pixel values are independent. Indeed, for $N$ independent trials, and in the limit of small false detection probability, the probability for a false signal classification scales as $P_{\text{false}, N}\propto N\cdot P_{\text{false},1}$. However, recovered pixel values are generally correlated. Therefore assuming $N_\text{pix}$ independent trials tends to underestimate the global significance of our detections.}

The pixel threshold values for maps with $N_{\rm side}=2$, and for a GW frequency of $f_{\rm GW}=3.95\,{\rm nHz}$ are reported in Fig.~\ref{fig:pixel_limits}. As expected, the lowest threshold values are found in the region of the sky with the highest density of pulsars. 

\subsection{Anisotropic signal-to-noise ratio}
Instead of defining a detection statistic based on the individual pixels, we can also use one that takes the full sky map into account. One such statistic is the anisotropic signal-to-noise ratio (SNR)~\cite{Pol:2022sjn}, which is given by the maximum likelihood ratio between an anisotropic and an isotropic sky:
\begin{equation}\label{eq:SNR}
    {\rm SNR} = \sqrt{2\ln\bigg[\frac{p(\boldsymbol{\rho}\vert\mathbf{\hat{P}})}{p(\boldsymbol{\rho}\vert \mathbf{P}_\text{iso})}\bigg]}\,,
\end{equation}
where $p$ is the likelihood function given in Eq.~\eqref{eq:likelikood}, $\mathbf{\hat{P}}$ is the recovered anisotropic sky map, and $\mathbf{P}_\text{iso}$ is the perfectly isotropic sky map. This detection statistics takes advantage of the full reconstructed sky, such that sources located in different regions of the sky can contribute to a classification as a detection (even if below the individual pixel thresholds discussed in the previous section).

Due to its global nature, this detection statistic is not suitable for assessing the significance of GW skies reconstructed using the radiometer approach. Indeed, the radiometer method only reconstructs the power of individual pixels while assuming that all the others do not contribute to the observed correlation pattern. Because of this, we will not use the SNR detection statistic in conjunction with the radiometer reconstruction method.

As for the pixel-based statistic, the SNR values are calibrated against null distributions to identify sky maps that give us a detection with a $p$-value $p<3\times 10^{-3}$ (corresponding to a $\sim3\sigma$ Gaussian-equivalent significance). 
%SNR null distributions and threshold for different values of $N_{\rm side}$ and $f_{\rm GW}$ are reported in Fig.~\ref{fig:snr_null} of App.~\ref{app:add_material}.

\subsection{$\chi^{2}$-Statistic}
Finally, we consider a third test statistic based on the ``distance'' between the reconstructed sky map and the ``typical'' recovered map for an underlying isotropic sky. Specifically, assuming that the uncertainties on the pulsar-pair correlations are normally distributed, the null distribution of the recovered pixel values also follows a multivariate normal distribution, $\mathcal{N}(\mathbf{\mu}, \mathbf{\Sigma})$, where $\mathbf{\mu}$ and $\mathbf{\Sigma}$ are the mean and covariance of the recovered pixel values for the sky maps of the null distribution. We can than calculate a probability distance estimator, the Mahalanobis distance~\cite{mahalanobis1936generalized}, for a specific sky map, $\mathbf{P}$, as
\begin{equation}
    D^{2}=(\mathbf{P}-\mu)^{T}\mathbf{\Sigma}^{-1}(\mathbf{P}-\mu).
\end{equation}

As for the previous test statistics, we compare the Mahalanobis distance for any recovered map with its null distribution. In this case the null distribution can be derived analytically, it follows a $\chi^{2}$ distribution with the number of degrees of freedom given by the number of pixels. Recovered skies that produce a Mahalanobis distance with a $p$-value smaller than $p<3\times 10^{-3}$ are considered as detections. 
 
Our application of the Mahalanobis distance as a detection statistic assumes that the null distribution is given by a multivariate normal distribution. While this is the case for both the radiometer pixel basis and the analytical map reconstruction, the positivity constraint that we impose during the numerical reconstruction changes the shape of the null distribution such that we can not use the Mahalanobis distance as an estimator for anisotropy detection in combination with the numerical approach.

\section{Results}\label{sec:results}
The main result of our analysis consists of an estimate of the anisotropy detection probability for different combinations of SMBHB population parameters. For each set of these parameters, $\vec\theta$, the anisotropy detection probability, $p_{\vec\theta}$, is estimated as 
\begin{equation}
    p_{\vec\theta}=\frac{N_{\vec{\theta},\rm det}}{N_{\rm tot}}\,,
\end{equation}
where $N_{\rm tot}=1500$ is the total number of mock GWB skies generated for each set of SMBHB population parameters, and $N_{\vec{\theta},\rm det}$ is the fraction of those skies that gave us an anisotropy detection.

\begin{figure*}[ht!]
    \centering
    \includegraphics[width=\textwidth]{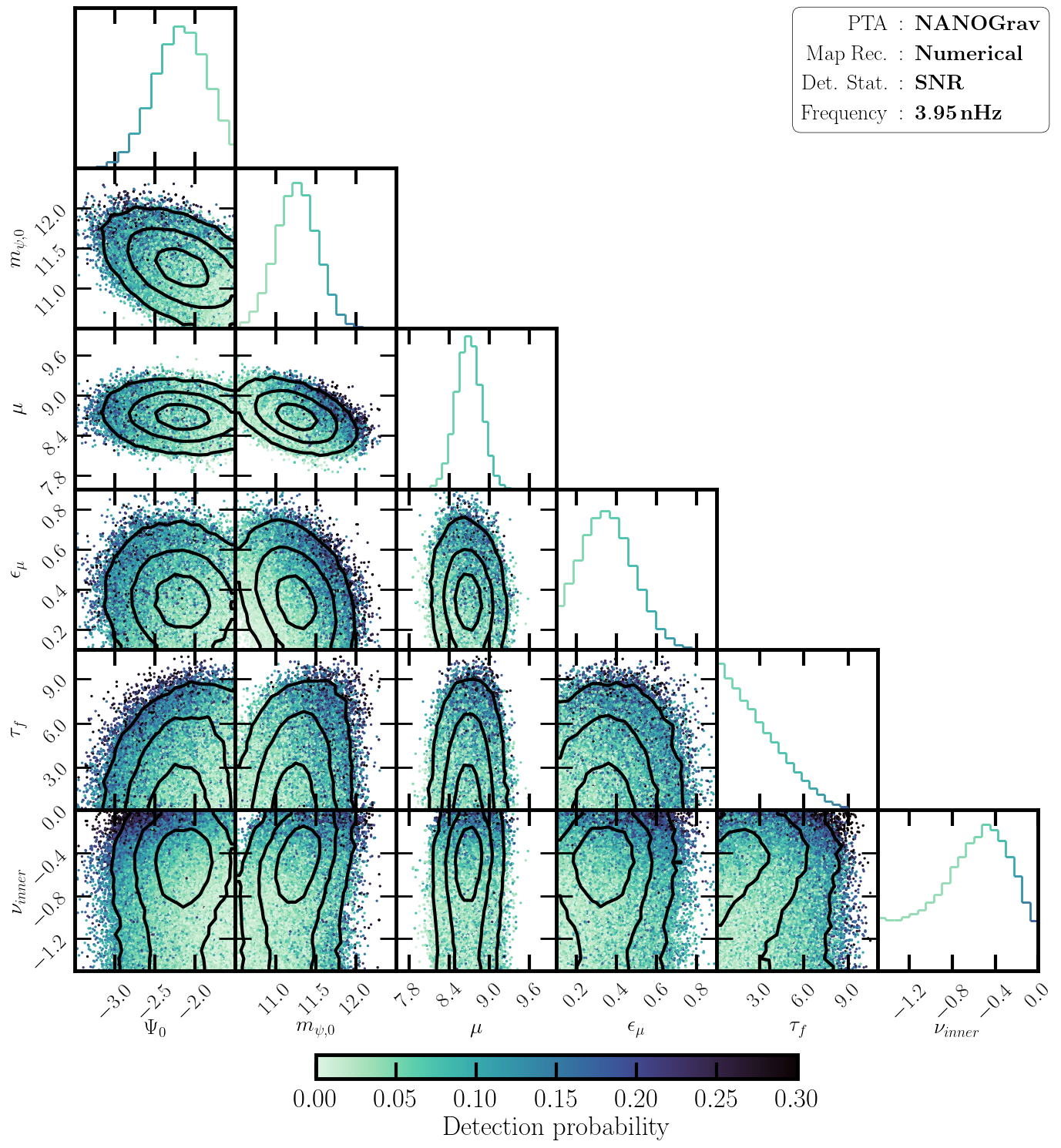}
    \caption{The off-diagonal panels show the 2D posterior distribution for the SMBHB population parameters derived in Ref.~\cite{ng+23_astro} by fitting the SMBHB signal to the observed GWB spectrum. The black contours represent the 68\%, 95\% and 99.7\% Bayesian credible regions for these posterior distributions, while the dots show the MCMC samples from which the posterior distributions are derived. The individual MCMC samples are colored according to their estimated anisotropy detection probability, $p_{\vec\theta}$. The diagonal panels show the marginalized 1D posterior distribution, the lines are colored according to the marginalized anisotropy detection probability.}
    \label{fig:corner}
\end{figure*}

In Fig.~\ref{fig:corner}, we report the posterior distributions of the SMBHB parameters used in this work. These distributions were derived in Ref.~\cite{ng+23_astro} by fitting the spectrum of SMBHB-sourced GWBs against the NANOGrav 15-year data. In this figure, the individual samples are colored according to their anisotropy detection probability, $p_{\vec{\theta}}$, for a NANOGrav-like PTA at $f_{\rm GW}=3.95\,{\rm nHz}$ (the most sensitive frequency bin). From this plot, we can identify some general trends in the detection probability values. The parameter that mostly impacts the detection probability is $\nu_{\rm inner}$, with larger values of this parameter leading to larger anisotropy detection probabilities (up to $p_{\vec\theta}\sim0.45$ for $\nu_{\rm inner}\sim0)$. This is consistent with the results of Ref.~\cite{Gardiner:2023zzr}, where it was shown that larger $\nu_{\rm inner}$ lead to a faster evolution of the binaries at small separation and fewer sources contributing to the background in the lowest frequency bins. Averaging over the SMBHB population parameters, we find an average detection probability of $\bar p_{\vec\theta}\simeq0.07$. This suggests that the null-detection for anisotropies reported in Ref.~\cite{ng+23_anis} is still perfectly compatible with an SMBHB interpretation of the GWB. 

The observed trends in detection probabilities suggest that anisotropy detection (or lack thereof) provides additional information about the SMBHB properties in addition to the ones already encoded in the background spectral properties. This extra information could be used in future analyses to break the degeneracy between SMBHB populations leading to similar GWB spectra. Specifically, we envision an analysis in which the information provided by a positive (or negative) anisotropy detection is included in the likelihood used in Monte Carlo explorations of the SMBHB parameter space.

\begin{figure}[t!]
    \centering
    \includegraphics[width=.48\textwidth]{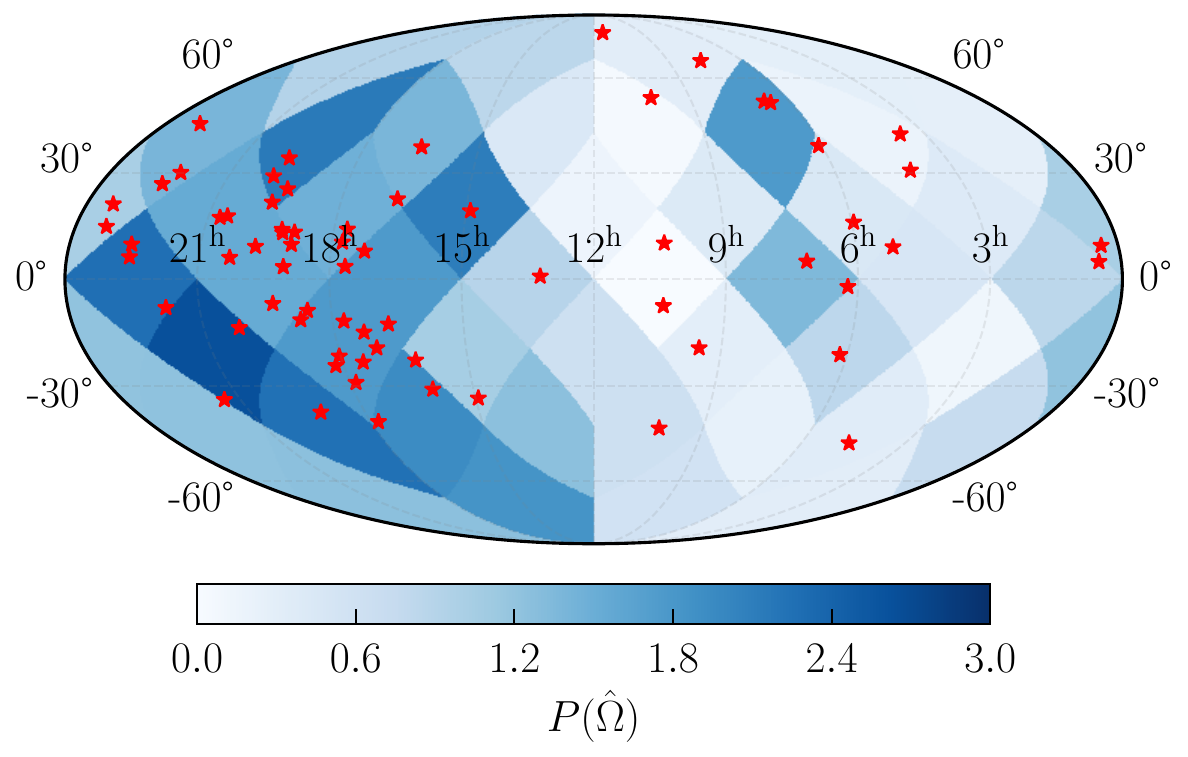}
    \caption{Average of the GWB skies (at $f_{\rm GW}=3.95\,{\rm Hz}$) that would lead to a detection of anisotropies in a PTA dataset with the noise properties of NANOGrav 15-year data set (using numerical map reconstruction and SNR-based detection statistic). The red stars indicate the position of the pulsars contained in the NANOGrav 15-year data set.}
    \label{fig:avg_sky}
\end{figure}

\begin{figure*}[t!]
    \centering
    \includegraphics[width=\textwidth]{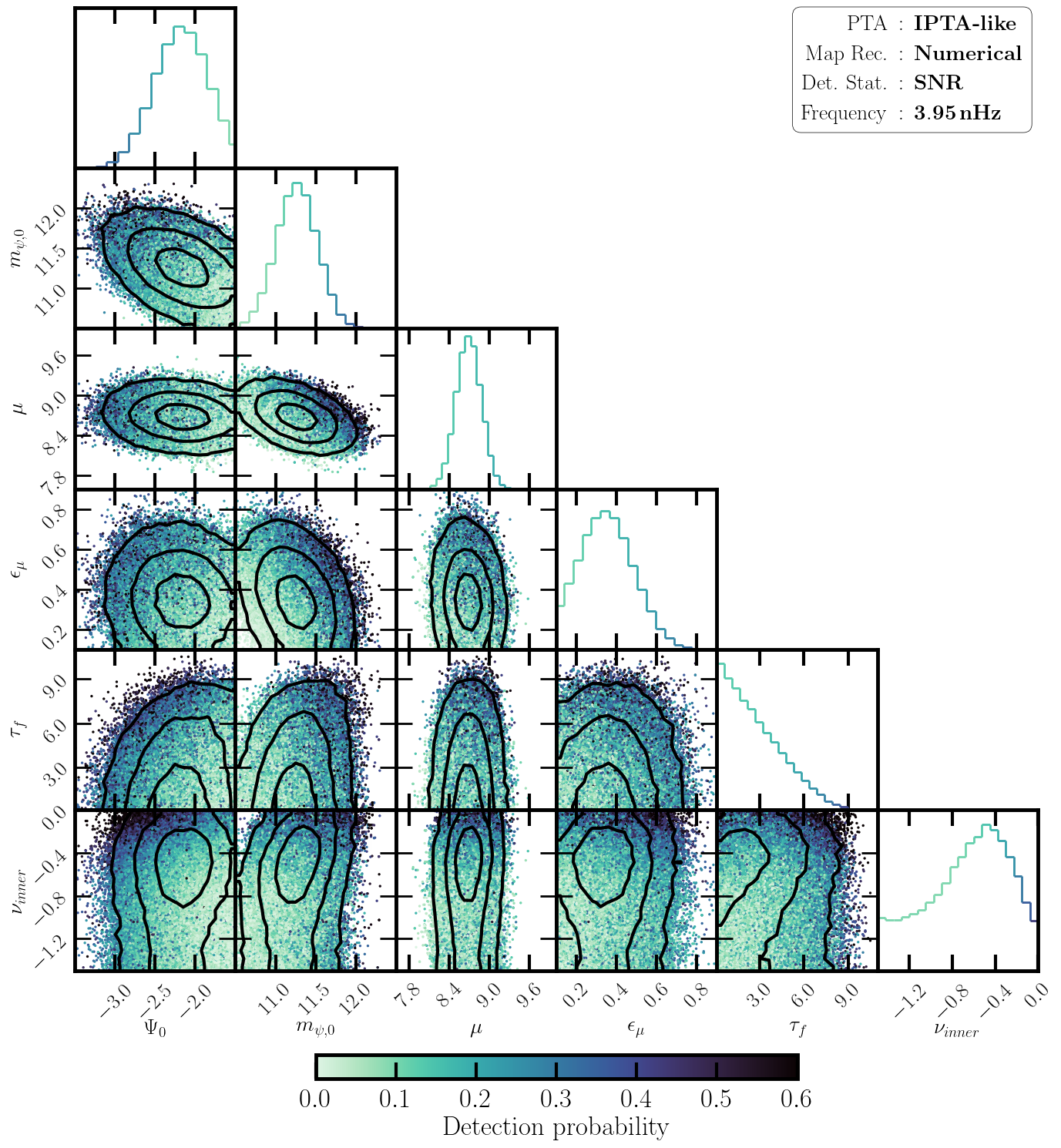}
    \caption{Same for Fig.~\ref{fig:corner} but for a PTA containing all the pulsars currently observed by EPTA+InPTA, MeerKAT, NANOGrav, and PPTA.}
    \label{fig:corner_ipta}
\end{figure*}

As expected, we find that most of the GW skies leading to a detection present a hot spot in the region of the sky in which we have the highest density of pulsars, and consequently a higher sensitivity to GWB anisotropies. This fact is clearly illustrated by Fig.~\ref{fig:avg_sky}, which reports the average of the detected GW skies, and shows how the typical detectable sky has more power in the region of the sky that is more densely populated by pulsars included in the data set.  This fact also suggests that searches for anisotropies will greatly benefit from the more uniform sky coverage that will be provided by the upcoming IPTA DR3. Estimates for anisotropy detection probabilities in an IPTA-like data set are reported in Fig.~\ref{fig:corner_ipta}. We find that, for this PTA, detection probabilities can reach up to $p_\theta\sim0.85$ for $\nu_{\rm inner}\sim0$, while the average detection probability grows to $\bar p_{\vec\theta}\sim0.16$. This result illustrates how future anisotropy searches could provide the first evidence in support of SMBHB as the origin of the background, or --in case of a null detection-- put this interpretation under pressure.

\begin{figure*}[t!]
    \centering
    \includegraphics[width=\textwidth]{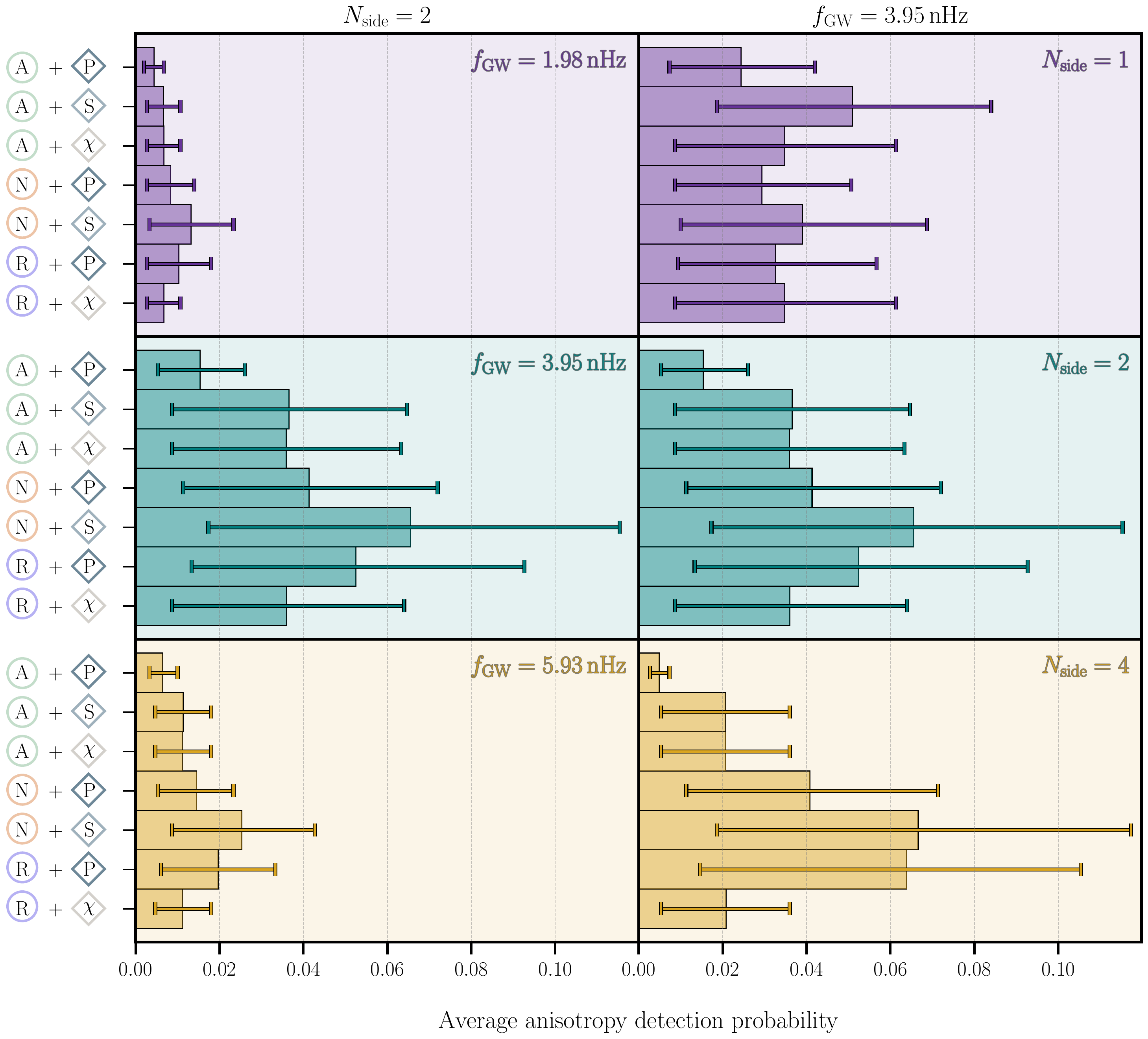}
    \caption{The left panel shows the average (histograms) and $68\%$ intervals (bars) for anisotropy detection probability (for $N_{\rm side}=2$) in the first three frequency bins and for different combinations of the map reconstruction method (numerical \protect\circled[numeric]{\textcolor{black}{N}}, analytical \protect\circled[analytic]{\textcolor{black}{A}}, or radiometer \protect\circled[radiometer]{\textcolor{black}{R}}) and detection statistic (SNR \protect\squared[snr]{\textcolor{black}{S}},pixel-based \protect\squared[pixel]{\textcolor{black}{P}}, or $\chi^2$-based statistic \protect\squared[chi]{{\textcolor{black}{$\chi$}}}). The right panel shows the same quantities for the second frequency bin and different values of $N_{\rm side}$.}
    \label{fig:prob_mean}
\end{figure*}

While in Fig.~\ref{fig:corner} and~\ref{fig:corner_ipta} we only report detection probabilities for the second frequency bin, we have also derived analogous results for the remaining lowest five frequency bins. We find that detection probabilities at other frequencies show similar trends in the SMBHB parameter space, even if with different overall normalizations. The left panel of Figure~\ref{fig:prob_mean} reports the average detection probabilities in the first three frequency bins, and shows how detection probabilities are the largest in the second (we have also checked detection probabilities at higher frequencies, finding that sensitivity to anisotropies keeps degrading going to higher frequencies). The second frequency bin results the most likely to detect GW anisotropies from SMBHB due to the balancing of competing effects: the level of anisotropies from SMBHBs is expected to increase at higher frequencies due to decreasing numbers of binaries contributing to the signal at those frequencies~\cite{Gardiner:2023zzr}; however, the red-tilt of the GWB spectrum causes the cross-correlation uncertainties to strongly increase at higher frequencies, limiting the sensitivity to anisotropies to the lower frequency range. The inclusion of pulsars with a timing baseline significantly shorter than 15 years introduces additional uncertainties in the first frequency bin, which leaves the second bin as the one most sensitive to searches for anisotropies.

While in Fig.~\ref{fig:corner} and Fig.~\ref{fig:corner_ipta} we only report the results derived with a combination of numerical map reconstruction and SNR-based detection statistic, we also derived these detection probabilities using all possible combinations of sky reconstruction techniques (discussed in Sec.~\ref{sec:search_strat}) and detection statistics (discussed in Sec.~\ref{sec:det_stat}). We find that for all these possible map reconstruction-detection statistic combinations we observe the same detection probability trends in the SMBHB parameter space, but with different overall normalization. Specifically, we find that numerical map reconstruction combined with the SNR-based detection statistics leads to the best detection prospects when coupled with a map resolution of $N_{\rm side}=2$ or $N_{\rm side}=4$ (see the right panel of Fig.~\ref{fig:prob_mean}). Except for the lower resolution maps ($N_{\rm side}=1)$, the analytical reconstruction tends to underperform the numerical reconstruction methods, while the radiometer basis delivers similar results.  This shows how, despite realistic SMBHB skies presenting more than a single bright pixel, the radiometer basis is still able to resolve them.

\begin{figure*}[ht!]
    \centering
    \includegraphics[width=\textwidth]{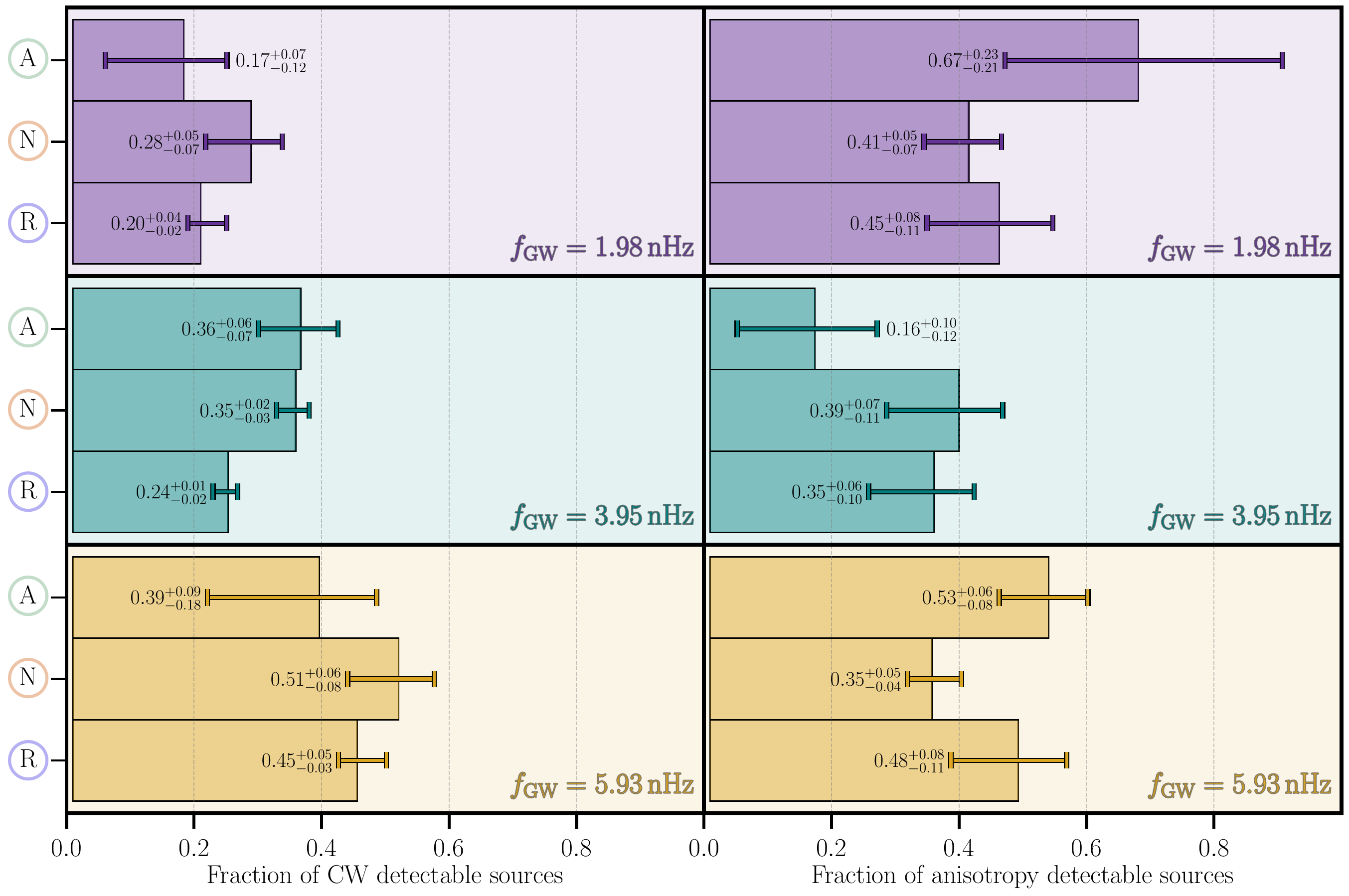}
    \caption{{\bf Left:} Fraction of anisotropy detections above the CW upper limits. {\bf Right:} Fraction of CW detectable skies that also led to an anisotropy detection.}
    \label{fig:cw_fraction}
\end{figure*}

Since most of the GW skies that lead to a positive detection for anisotropies are characterized by a loud binary, it is interesting to consider the interplay with continuous wave (CW) searches. By comparing the CW upper limits provided in Ref.~\cite{ng+23_cw} with the characteristic strain of all binaries in our mock GW skies, we find that roughly $35\%$ of the anisotropy detections in the second frequency bin would be accompanied by a positive detection in CW searches at the $2\sigma$-level (see the left panel of Fig.~\ref{fig:cw_fraction}). Similarly, we find that around $40\%$ of the skies with a CW-detectable binary in the second frequency bin would also lead to a positive anisotropy detection (see the right panel of Fig.~\ref{fig:cw_fraction}). Of course, these results are only rough estimates based on marginalized CW upper limits. A more detailed analysis would require performing a full CW search on the SMBHB populations used to derive the mock GW skies. We leave this analysis to future studies~\cite{emiko}.

Finally, we want to emphasize that in deriving our results we ignored the impact that interference between the binaries signals can have on the pulsars cross correlations~\cite{Allen:2022dzg, Allen:2022ksj}. This effect --which is a combination of what is commonly known as \emph{pulsar variance} and \emph{cosmic variance}-- can induce significant deviations from the expected cross-correlation values given in Eq.~\eqref{eq:theo_corr} and make anisotropy detection more challenging. The precise impact of these effects, together with possible mitigation strategies, will be discussed in future work~\cite{konstandin:2024}. However, because of this approximation, the detection probabilities reported herein should be regarded as optimistic estimates, contingent upon improvements in detection methods to account for such interference effects.

\section{Discussion}
In this work, we developed a pipeline to perform realistic anisotropy searches on GWBs produced by synthetic SMBHB populations. We then used this framework to study the prospects of anisotropy detection by current and future PTAs, under the assumption that the observed GWB is produced by a population of SMBHB. The main results of our analysis are the following:
\begin{enumerate}
    \item Based on our analysis, the lack of evidence for GWB anisotropies in a PTA with the noise characteristics of the NANOGrav 15-year data set is not in tension with assuming SMBHBs as a source of the observed background. Indeed, we find that for a SMBHB-generated GWB there was only a $\sim7\%$ probability of detecting anisotropies in such a data set. 
    \item We estimate that the probability of detecting GWB anisotropies in a SMBHB-generated background will increase to $\sim16\%$ for the upcoming IPTA DR3 data set. This indicates that, as we keep improving the sensitivity of our observations, a null anisotropy detection will start to put pressure on an SMBHB interpretation of the background. 
    \item We find that the anisotropy detection probability strongly depends on the properties of the SMBHB population. For example, while the average detection probability is around $\sim7\%$ for a PTA with the noise properties of NANOGrav 15-year data set, probabilities can reach up to $\sim45\%$ for SMBHB populations where the phenomenological hardening processes speed up at smaller separations (i.e., when $\nu_{\rm inner}\sim0)$. This indicate that an anisotropy detection (or lack of thereof) can convey information about the SMBHB population properties, and break the degeneracy between SMBHB populations with similar GWB spectral properties.
    \item Among the considered reconstruction methods and detection statistics, the combination that performs the best uses a numerical map reconstruction method and an SNR-based detection statistic. 
    \item We find that only $\sim40\%$ of the GWB skies leading to detection of anisotropies contain a binary with a CW signal above current CW upper limits. This emphasizes the complementarity between CW and anisotropy searches in the effort to detect the first direct evidence for SMBHB being the source of the GWB. 
\end{enumerate}

\section*{Acknowledgments}
We would like to thank Emiko C. Gardiner, Nihan Pol, Luke Zoltan Kelley, Tristan Smith, Stephen R. Taylor and the entire NANOGrav collaboration for useful discussions and comments on a preliminary form of this work. The work of AM and AL was supported by the Deutsche Forschungsgemeinschaft under Germany’s Excellence Strategy - EXC 2121 Quantum Universe - 390833306. KAG acknowledges support from an NSF CAREER \#2146016. This work used the Maxwell computational resources operated at Deutsches Elektronen-Synchrotron DESY, Hamburg (Germany).

\bibliographystyle{apsrev4-1}
\bibliography{references}

\end{document}